\documentclass{webofc}
\usepackage[varg]{txfonts}   
\usepackage{braket}
\usepackage{multirow}
\usepackage{multicol}
\usepackage{subcaption,booktabs}
\begin{document}
\title{Gravitational coupling of QED and QCD: 3- and 4- point functions in momentum space}

%
%

\author{    
    \firstname{Matteo Maria}        \lastname{Maglio}\inst{1}\fnsep\thanks{\email{maglio@thphys.uni-heidelberg.de}} 
\and
    \firstname{Riccardo} 
    \lastname{Tommasi}\inst{2}\fnsep\thanks{\email{riccardo.tommasi@unisalento.it}}
}

\institute{
    Institute for Theoretical Physics, University of Heidelberg, Germany
\and
    Dipartimento Matematica e Fisica "Ennio de Giorgi", Università del Salento, Lecce, Italy
          }

\abstract{%
  Conformal symmetry has important consequences for strong interactions at short distances and provides powerful tools for practical calculations. Even if the Lagrangians of Quantum Chromodynamics (QCD) and Electrodynamics (QED) are invariant under conformal transformations, this symmetry is broken by quantum corrections. The signature of the symmetry breaking is encoded in the presence of massless poles in correlators involving stress-energy tensors. We present a general study of the correlation functions $\braket{T J J}$ and $\braket{T T J J}$ of conformal field theory (CFT) in the flat background limit in momentum space, following a reconstruction method for tensor correlators. Furthermore, our analysis also focuses on studying the dimensional degeneracies of the tensor structures related to these correlators.
}
\maketitle
\section*{Introduction}

Correlation functions play a significant role in conformal field theories (CFT), and their functional form can easily be obtained by methods in coordinate space rather than in momentum space. Although these methods are very powerful, they apply when the correlation functions are at separate points. On the other hand, anomalies occur in coordinate space at short distances and thus with configurations of a correlator in which some points coalesce. Then the coordinate space approach provides limited information on the origin of the anomaly, except for telling us that its origin is a short-distance effect. 
One of the main reasons for studying CFT correlation functions directly in momentum space is to see the effects of anomalies more directly. 
When a theory is classically conformal invariant, then as a consequence, its energy momentum tensor will have a null trace. If the quantum theory has an anomaly, then the trace of the expectation value of the energy momentum tensor in a metric ($g_{\mu\nu}$) background develops a non-zero value. 
This phenomenon is essentially related to divergences in the quantum theory that break the conformal invariance once the theory is renormalized.
In particular, two counterterms are needed in $d=4$ in the renormalization of correlators with multiple stress-energy tensors, responsible for the generation of the anomaly: $E$ and $C^2$, the Euler-Poincar\`e density and the Weyl tensor squared respectively, which find application in the $\braket{TTT\dots}$ ($n$-graviton) vertex \cite{Coriano:2017mux, Coriano:2021nvn,Coriano:2022vrz}. For other correlators, such as the $\braket{TJJ}$, the renormalization of the vector 2-point function $\braket{JJ}$ is sufficient to generate a finite correlator, which is reflected in the $F^2$ term of the anomaly functional, with $F$ the field strength of the gauge field (see \cite{Coriano:2018bbe,Coriano:2018zdo} for related studies). 

\section*{Anomalies as light cone processes}

The perturbative realization of CFT correlators allows us to handle far more simplified expressions proceeding with analyzing an ordinary Feynman expansion. The analysis in momentum space of such correlators provides additional information on the emergence of the conformal anomaly. Indeed, the appearance of the anomaly can be described by the emergence of massless effective scalar degrees of freedom in the 3-point functions containing insertions of stress energy tensors, which can be interpreted as light-cone interactions. As discussed in \cite{Giannotti:2008cv,Armillis:2010qk,Armillis:2010ru,Coriano:2020ees,Coriano:2012wp}, this phenomenon points towards an interpretation of the origin of the conformal anomaly as mediated by correlated pairs of fermions/scalars, as emerging from the spectral representation of a given perturbative correlator. 
Such interactions also play a role in the context of Weyl semimetals, with the paired electrons (representing the massless pole) interacting with the lattice of such materials \cite{Chernodub:2019tsx,Chernodub:2017jcp}. It is worth mentioning that both conformal and chiral anomalies play a key role in this phenomenon \cite{Chernodub:2021nff}.
These interactions are associated with renormalization and are not related to specific parametrization of the tensor correlators. The proof can be illustrated more quickly in the case of the $TJJ$ correlator in QED. A similar analysis can be performed for QCD, even if it is more involved. 
\section*{Decomposition of correlators}

We briefly review the method to decompose any $n$-point functions involving tensorial operators, first presented for the case of the three-point functions \cite{bms}. This method is based on the reconstruction of the full $n$-point function involving stress-energy tensors, currents, and scalar operators starting from the expressions of transverse and traceless part only. We will present the decomposition of $TJJ$ and $TTJJ$ correlation functions directly in momentum space. Defining the projectors
\begin{equation}
    \pi^{\mu}_{\alpha}=\delta^{\mu}_\alpha-\frac{p^\mu p_\alpha}{p^2}, \qquad \Pi^{\mu\nu}_{\alpha\beta}=\pi^{(\mu}_\alpha\pi^{\nu)}_\beta-\frac{1}{d-1}\pi^{\mu\nu}\pi_{\alpha\beta},
    \end{equation}
with the properties 
\begin{equation}
p_{i\mu_i}\,\pi^{\mu_i\nu_i}(p_i)=0,\quad p_{i\mu_i}\,\Pi^{\mu_i\nu_i}_{\alpha_i\beta_i}(p_i)=0,\qquad \delta_{\mu_i\nu_i}\,\Pi^{\mu_i\nu_i}_{\alpha_i\beta_i}(p_i)=0,
\end{equation}
we consider the decomposition of the energy-momentum tensor $T^{\mu\nu}$  and the current $J^\mu$ as
\begin{align}
     &j^{\mu}(p)=\pi^{\mu}_{\alpha}(p)\,J^{\alpha}(p), &&
    j_{loc}^{\mu}(p)=\frac{p^{\mu}p_\alpha}{p^2}J^{\alpha}(p),\notag\\
  & t^{\mu\nu}(p)=\Pi^{\mu\nu}_{\alpha\beta}(p)\,T^{\alpha\beta}(p), &&
    t_{loc}^{\mu\nu}(p)=\left(I^{\mu\nu}_{\alpha\beta}+\frac{1}{d-1}\pi^{\mu\nu}\delta_{\alpha\beta}\right)T^{\alpha\beta}(p), \label{proj}\\
   &&&\hspace{-4.9cm} I^{\mu\nu}_{\alpha\beta}=\frac{p_\beta}{p^2}\left[2p^{(\mu}\delta^{\nu)}_\alpha-\frac{p_\alpha}{d-1}\left(\delta^{\mu\nu}+(d-2)\frac{p^\mu p^\nu}{p^2}\right)\right]\notag.
\end{align}
The decomposition of the operators $T^{\mu\nu}=t^{\mu\nu}_{loc}+t^{\mu\nu}$ and $J^\mu=j^\mu+j_{loc}^\mu$, allows to split any correlation function into a sum of correlators containing $j^\mu$, $j^\mu_{loc}$, $t^{\mu \nu}$ and $t^{\mu \nu}_{loc}$. However, as shown in \cite{bms, Bzowski:2017poo,Coriano:2020ees}, by using the conservation Ward Identities, that relate $n$-point functions to lower point, it is possible to completely fix the longitudinal parts, i.e. those terms containing at least one $t_{loc}$ or $j_{loc}$. Therefore, the only term to be studied in order to reconstruct the entire correlator is the transverse traceless part consisting only of operators $t^{\mu \nu}$ and $j^\mu$. \\
The transverse traceless part, as we will show, can be expressed in a number of minimal tensor structures and form factors. Furthermore, due to the presence of dimensional tensor degeneracies, the number of independent tensor structures contributing to the decomposition can be properly reduced.

\section*{$TJJ$ reconstruction}
In this section we give the decomposition of the correlator $TJJ$, and in particular, by using \eqref{proj}, we obtain
\begin{align*}
    \braket{T^{\mu_1\nu_1}\,J^{\mu_2}\,J^{\mu_3}}&=\braket{t^{\mu_1\nu_1}\,j^{\mu_2}\,j^{\mu_3}}+\braket{T^{\mu_1\nu_1}\,J^{\mu_2}\,j_{loc}^{\mu_3}}+\braket{T^{\mu_1\nu_1}\,j_{loc}^{\mu_2}\,J^{\mu_3}}+\braket{t_{loc}^{\mu_1\nu_1}\,J^{\mu_2}\,J^{\mu_3}}\notag\\
    &\quad-\braket{T^{\mu_1\nu_1}\,j_{loc}^{\mu_2}\,j_{loc}^{\mu_3}}-\braket{t_{loc}^{\mu_1\nu_1}\,j_{loc}^{\mu_2}\,J^{\mu_3}}- \braket{t_{loc}^{\mu_1\nu_1}\,J^{\mu_2}\,j_{loc}^{\mu_3}}+\braket{t_{loc}^{\mu_1\nu_1}\,j_{loc}^{\mu_2}\,j_{loc}^{\mu_3}}.
    \end{align*}
It is important to emphasise that all the terms except the first one can be rewritten as two-point functions via Ward identities. The explicit form of the transverse traceless part $\braket{t^{\mu_1\nu_1}\,j^{\mu_2}\,j^{\mu_3}}$ is 
	\begin{equation} 
	\braket{ t^{\mu_1 \nu_1} (p_1)j^{\mu_2} (p_2) j^{\mu_3} (p_3) } =\Pi^{\mu_1 \nu_1}_{\alpha_1 \beta_1} (p_1) \pi^{\mu_2}_{\alpha_2} (p_2) \pi^{\mu_3}_{\alpha_3} (p_3) X^{\alpha_1 \beta_1 \alpha_2\alpha_3},\label{tjj}
\end{equation}
where $X^{\alpha_1\dots\alpha_3}$ is a general rank four tensor built by products of metric tensors and momenta with the appropriate choice of indices. As a consequence of the projectors in \eqref{tjj}, $X^{\alpha_1\dots\alpha_3}$ can not be constructed by using $\delta^{\alpha_1\beta_1}$, nor by $p_i^{\alpha_i}$, $i=1,\dots,3$. In addition, the conservation of the total momentum
\begin{equation} 
	p_1^{\alpha_i} + p_2^{\alpha_i} + p_3^{\alpha_i}= 0 ,
\end{equation}
allows selecting for each index $\alpha_i$ a pair of momenta to be used in the general construction of $X$. The choice of the independent momenta of the expansion can be different for each set of contracted tensor indices. One can choose   
\begin{equation}
	\begin{split}
		&\{\alpha_1,\beta_1\}\leftrightarrow p_1,p_2,\\
		&\{\alpha_2\}\leftrightarrow p_2,p_3\,,\\
		&\{\alpha_3\}\leftrightarrow p_3,p_1\,,
	\end{split}\label{choicemom}
\end{equation}
as basis of the expansion for each pair of indices shown above. The linear dependence of one momentum, for instance $p_3$, which we will impose at a later stage, is not in contradiction with this choice, which allows to reduce the number of form factors, due to the presence of a single $t$ projector for each external momentum. For which concerns the metric delta the only non vanishing terms appearing in $X^{\alpha_1\dots\alpha_3}$ are
\begin{align}
	\delta^{\alpha_1\alpha_2},\ \delta^{\alpha_1\alpha_3},\ \delta^{\alpha_2\alpha_3},\label{choicemetr}
\end{align}
together with the terms obtained by the exchange $\alpha_1\leftrightarrow\beta_1$. To construct the transverse traceless part, we must use these tensors to build all possible four rank tensors. Still, we must keep in mind that, due to symmetries of the correlator, form factors associated with structures linked by a  $2 \leftrightarrow 3$ transformation are dependent. Then the transverse traceless part is written as
  \begin{align} \label{tjjdecomposition}
   & \langle t^{\mu_1\nu_1}(p_1)j^{\mu_2}(p_2)j^{\mu_3}(p_3)\rangle =\notag\\
    &=
    {\Pi}^{\mu_1\nu_1}_{\alpha_1\beta_1}(p_1)\,{\pi}^{\mu_2}_{\alpha_2}(p_2)\,{\pi}^{\mu_3}_{\alpha_3}(p_3)\,
    \big( A_1\ p_2^{\alpha_1}p_2^{\beta_1}p_3^{\alpha_2}p_1^{\alpha_3} + 
    A_2\ \delta^{\alpha_2\alpha_3} p_2^{\alpha_1}p_2^{\beta_1} \notag \\
   &\quad+A_3\ \delta^{\alpha_1\alpha_2}p_2^{\beta_1}p_1^{\alpha_3} + 
    A_3(p_2\leftrightarrow p_3)\delta^{\alpha_1\alpha_3}p_2^{\beta_1}p_3^{\alpha_2}
    + A_4\  \delta^{\alpha_1\alpha_3}\delta^{\alpha_2\beta_1}\big),
         \end{align}  
It is worth mentioning that the form factors $A_i$ are fixed by the conformal invariance and in particular by the CWIs
    \begin{equation*}
        K^\kappa\braket{T^{\mu_1\nu_1}\,J^{\mu_2}\,J^{\mu_3}}=0,\qquad D\braket{T^{\mu_1\nu_1}\,J^{\mu_2}\,J^{\mu_3}}=0,
    \end{equation*}
where $D$ and $K^\kappa$ are the dilatation and the special conformal operators respectively \cite{Coriano:2013jba, bms, Bzowski:2015pba, Coriano:2019sth}  

\section*{$TJJ$ perturbative realization}
\begin{figure}[h!]
    \centering
    \includegraphics[scale=.18]{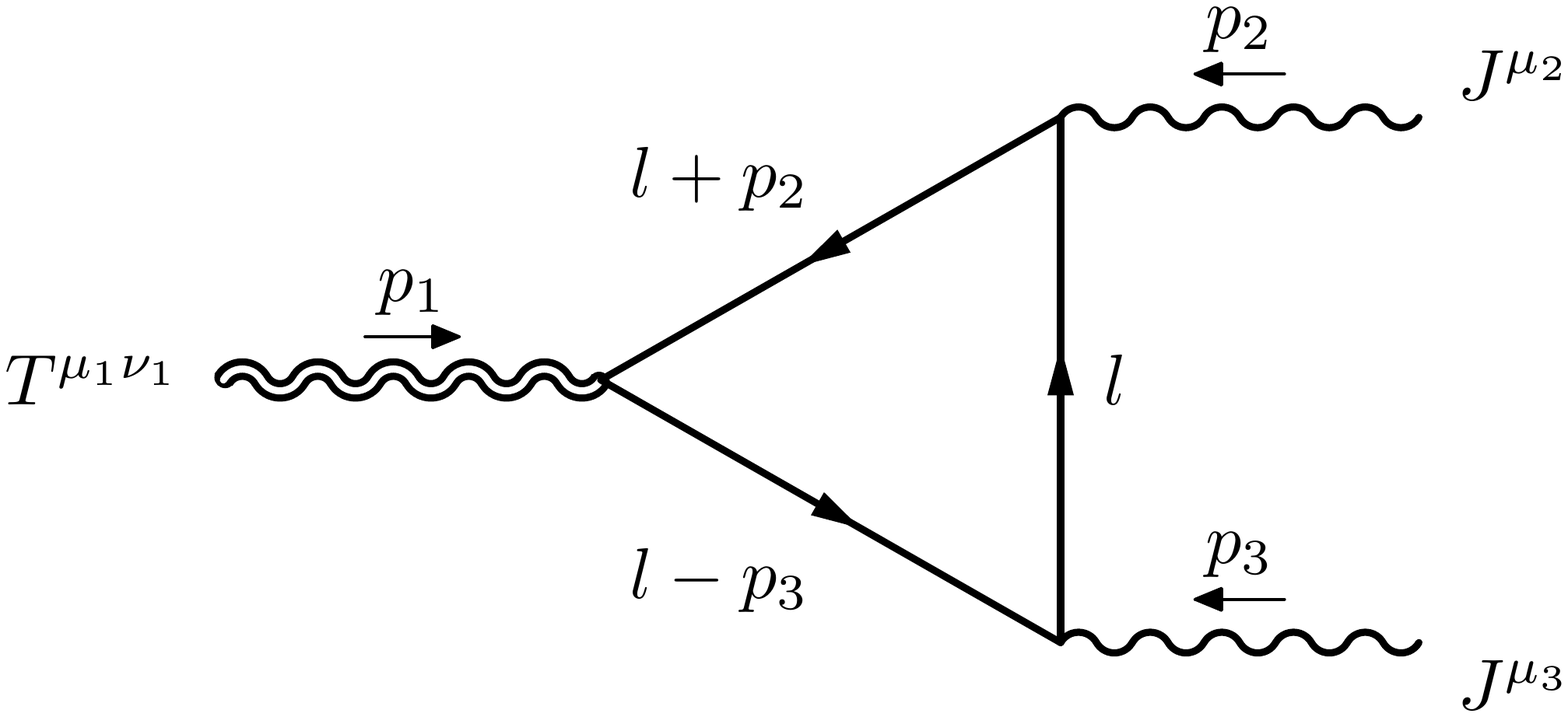}\,\raisebox{0.85cm}{$+$}\,
    \includegraphics[scale=.18]{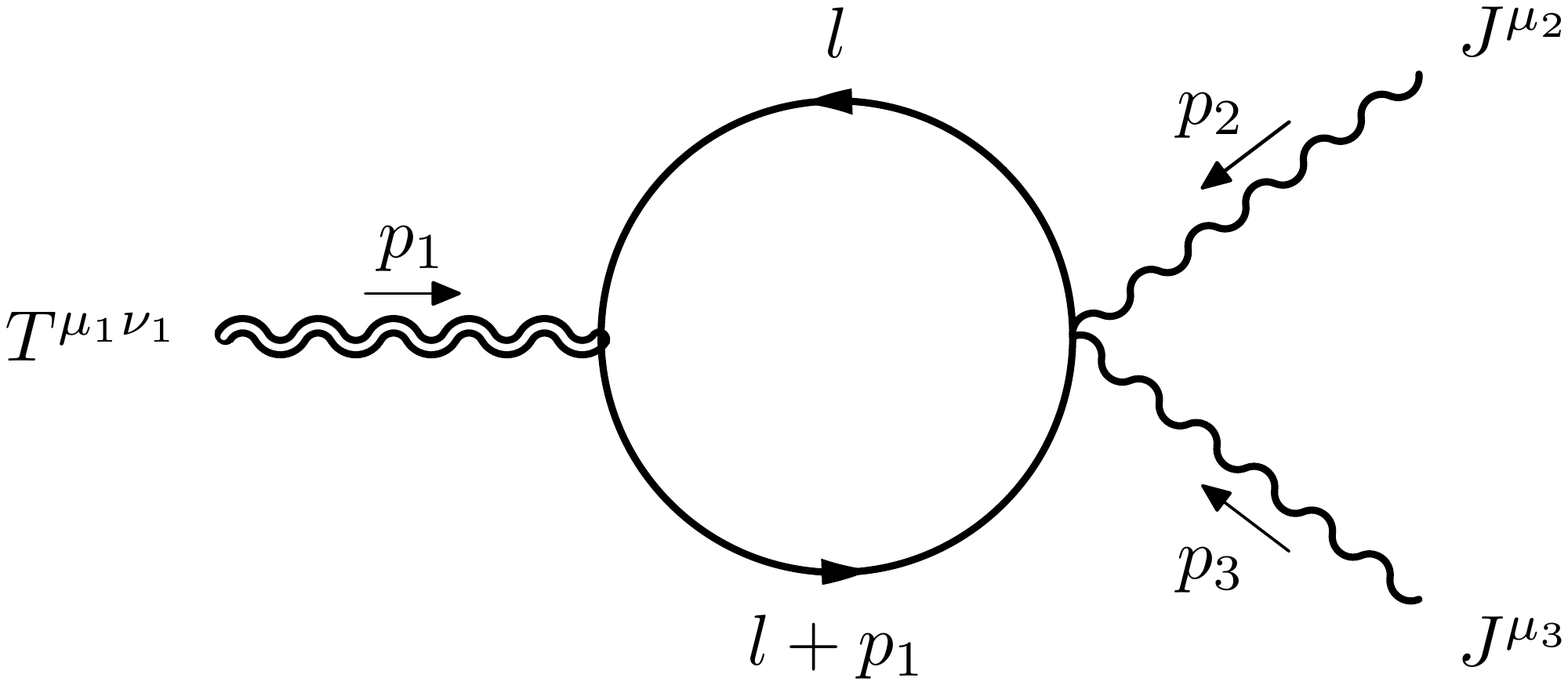}\,\raisebox{0.8cm}{$+ \cdots$}  
    \caption{One-loop diagrams for the $TJJ$ in QED. \label{fig1}}  
\end{figure}

A perturbative calculation of the $TJJ$ correlator has been performed in two different realization of QED, namely with complex scalars and fermions \cite{Coriano:2018bbe}. The actions considered are given by
   \begin{align} 
   S_{s} &=\int d^d x \, \sqrt{-g} \, \left( \, \partial^\mu \phi^\dagger \, \partial_\mu \phi + ie \, A^\mu \, (\partial_\mu \phi^\dagger \, \phi - \phi^\dagger \, \partial_\mu \phi)\, \, + \, \,e^2 A^\mu A_\mu \, \phi^\dagger \phi \, \, + \, \, \chi R \, \phi^\dagger \phi \right),\\
   S_{f} &= \int d^d x \, V \, \left(\, {i \over 2} \, ( \, \bar \psi \, \gamma^\lambda \, {\partial}_\lambda \psi \, -  \, \partial_\gamma \bar \psi \, \gamma^\lambda \, \psi)\, \, - \, \, e \, \bar \psi \, \gamma^\lambda A_\lambda \, \psi \, \, - \, \, {i \over 4} \, \omega_{\mu a b } \, V^\mu_c\, \bar \psi \gamma^{abc} \psi \, \right)
            \end{align}
where $V^\mu_a$ is the vielbein and $\omega_{\mu a b}$ is the spin connection. 
The correlator $TJJ$ is obtained from the sum of the Feynmann diagrams that can be constructed, and as an example we show some of them in Fig. \ref{fig1}. 

 From the calculation of the Feynman diagrams, the form factors in the decomposition can be written explicitly in terms of master integrals $B_0$ and $C_0$. In $d=4$, this correlator manifests UV divergences that can be renormalized by adding the gauge and Weyl invariant counterterm
    \begin{equation}
    S_{ct}=-\frac{c}{\epsilon} \, \int d^d x\sqrt{-g} \ F_{\mu \nu} F^{\mu \nu}\, .
    \end{equation}
    The renormalization procedure of the correlator induces a breaking of the conformal invariance, manifested in the presence of an anomaly pole. The correlator, which was classically conformal invariant, acquires, after renormalization, a trace anomaly contribution. 
    The effective action related to the anomaly contribution at the first order in the perturbation $h$ is 
    \begin{equation}
        S_{pole} = -\frac{e^2}{36 \pi^2} \, \int d^4 x \, d^4 y \, \, \Big(\square h (x) - \partial_\mu \partial_\nu h^{\mu \nu} (x) \Big) \, \square^{-1}_{xy} \, F_{\alpha \beta} (x)  F^{\alpha \beta} (y).
    \end{equation}

\section*{Dimensional dependent degeneracies}
The number of independent tensorial structures corresponds to the number of independent form factors that characterize the transverse traceless part. At first sight, it seems evident that different tensorial sectors (i.e. terms involving only metric $\delta$ and terms involving $\delta$ and momenta $p_i$) are not connected under symmetry transformations, and we can consider each sector individually when counting the number of independent form factors. \\
However, this is the case only when we work in general dimension $d$. If we are interested in a specific dimension, there may be relations that relate to terms among different sectors. One way to analyze such identities is by using Lovelock's double antisymmetrization method \cite{lovelock, edgar}. \\
The main point of this analysis is to consider a $(k,l)$-rank tensor $S$ in $d$ dimensions, with $k,l < d$. Then the relation
\begin{equation} \label{lovelock1}
	S_{[\alpha_1 \dots \alpha_k}^{\quad \beta_1 \dots \beta_l} \, \, \delta_{\alpha_{k+1} }^{\beta_{l+1} } \, \cdots \, \delta_{\alpha_{k+m} ]}^{\beta_{l+m}} = 0 \qquad \qquad \text{for } \, \, k+m>d \, \, ,
\end{equation} 
is trivially satisfied, and further contractions of these identities may lead to non trivial relations. \\
This result can be reformulated in a different way \cite{lovelock}. Let $T$ be an anti-symmetric traceless $(k,l)$-rank tensor, then the Lovelock's theorem states that
\begin{equation} \label{lovelock2}
	T_{[\alpha_1 \dots \alpha_k}^{\qquad [\beta_1 \dots \beta_l}  \, \, \delta_{\alpha_{k+1} }^{\beta_{l+1} } \, \cdots \, \delta_{\alpha_{k+m} ]}^{\beta_{l+m}]} = 0, \qquad \qquad \text{for } \, \, m \geq d+1-(k+l) \, \, .
\end{equation}
This relation may seem less intuitive, but it is the same principle of \eqref{lovelock1}. Since the number of indices exceeds the dimensionality, two indices will be either equal on the same line or contracted. In the latter case, being the tensor traceless, the identity will be satisfied. Lovelock's theorem ensures that any contraction of \eqref{lovelock2} results only in trivial identities. \\
This theorem provides evidence for tensor identities based on the dimension considered. Indeed, given the dimension $d$, we build an antisymmetric traceless $(k,l)$ rank tensor. We obtain a family of tensor identities using a double antisymmetrization in the form of \eqref{lovelock2}. For example, we could construct a Riemann-like tensor using $p_1$ and $p_2$ 
\begin{equation} 
	R_{\alpha_1 \alpha_2}^{\quad \beta_1 \beta_2} = {p_1}_{[ \alpha_1} {p_2}_{\alpha_2 ]} \, \, {p_1}^{[ \beta_1} {p_2}^{\beta_2 ]} \, \, .
\end{equation}
The antisymmetrization of this tensor with two metric tensors defines the traceless tensor $W$
\begin{equation}
	W_{\alpha_1 \alpha_2}^{\quad \beta_1 \beta_2}=R_{\alpha_3 \alpha_4}^{\quad [\alpha_3 \alpha_4}\delta_{\alpha_1}^{\beta_1}\delta_{\alpha_2}^{\beta_2]}
\end{equation}
that has the same properties as the Weyl tensor. In $d\le3$ this tensor vanishes identically, i.e.
\begin{equation}
	W_{\alpha_1 \alpha_2}^{\quad \beta_1 \beta_2}=0,\quad d\le3,
\end{equation}
as specific case of \eqref{lovelock2} with $k=l=2$, that in $d=3$ imply $m=0$ as pointed out in \cite{Bzowski:2017poo}. \\
Then there are possible values of $(k,l)$ depending on the specific dimensions that produce different identities on tensor structures, and they may reduce the number of independent form factors. 

\section*{$TTJJ$ reconstruction and perturbative realization}
The general decomposition presented can be extended in studying higher point functions. In the case of $\braket{TTJJ}$, the conservation of the total momentum is not as strict as the three-point case, and the symmetry of the correlator is now twofold, having both $(1\leftrightarrow2)$ and $(3\leftrightarrow4)$ permutation symmetry to be considered. 
The choice of the independent momenta in the transverse traceless part, as we did in \eqref{choicemom}, is similar for the four point function but with the association of two independent moments for each index, as pointed out in \cite{Coriano:2021nvn,Coriano:2019nkw}. 
More details to obtain the decomposition into minimal independent form factors will be presented in \cite{ttjj}. \\
The perturbative realization of the $\braket{TTJJ}$ can be obtained analogously to the case of the three-point function by calculating the corresponding Feynman diagrams.           
In both cases, when $d=4$, the renormalization procedure, due to the presence of divergences, breaks the conformal invariance, which is reflected in anomaly massless poles. We will present the form of the anomaly effective action for the $\braket{TTJJ}$ correlator studying its implications \cite{ttjj}. In addition, a detailed study on tensor degeneracies for the $\braket{TTJJ}$ decomposition will be investigated.

\section*{Conclusions}
We have briefly presented the general method to construct 3- correlation functions using the decomposition into a longitudinal part and a transverse traceless one. This method can be extended straightforwardly to the case of 4-point functions. The transverse traceless part of the correlator can be written in terms of independent form factors and tensor structures. The number of independent form factors is related to the number of independent tensorial structures that depends on the specific dimension chosen. Indeed, we have illustrated the case where tensor identities have to be considered in specific dimensions. These constraints change the number of independent form factors in the specific dimensions. Finally, in $d=4$, we have mentioned how the renormalization procedure of correlators involving energy momentum tensors leads to the presence of anomalous massless poles and the identification of the anomaly effective action for the $TJJ$. The tensor identities presented in this paper can play a crucial role in identifying the anomalous part of the $TTJJ$ in $d=4$. 

\section*{Acknowledgement}
M. M. M. is supported by the European Research Council (ERC) under the European Union as Horizon 2020 research and innovation program (grant agreement No818066) and by Deutsche Forschungsgemeinschaft (DFG, German Research Foundation) under Germany’s Excellence Strategy EXC-2181/1 - 390900948 (the Heidelberg STRUCTURES Cluster of Excellence). The work of R.T. is supported by INFN iniziativa specifica QFT-HEP. 
%
\bibliography{PoSbiblio}

\begin{thebibliography}{22}

\bibitem{Coriano:2017mux}
C.~Coriano, M.M. Maglio, E.~Mottola, Nucl. Phys. B \textbf{942}, 303 (2019),
  \texttt{1703.08860}

\bibitem{Coriano:2021nvn}
C.~Corian\`o, M.M. Maglio, D.~Theofilopoulos, Eur. Phys. J. C \textbf{81}, 740
  (2021), \texttt{2103.13957}

\bibitem{Coriano:2022vrz}
C.~Corian\`o, M.~Cret\`\i{}, S.~Lionetti, M.M. Maglio, R.~Tommasi (2022),
  \texttt{2205.03535}

\bibitem{Coriano:2018bbe}
C.~Corian\`o, M.M. Maglio, Nucl. Phys. B \textbf{938}, 440 (2019),
  \texttt{1802.07675}

\bibitem{Coriano:2018zdo}
C.~Corian\`o, M.M. Maglio, Phys. Lett. B \textbf{781}, 283 (2018),
  \texttt{1802.01501}

\bibitem{Giannotti:2008cv}
M.~Giannotti, E.~Mottola, Phys. Rev. D \textbf{79}, 045014 (2009),
  \texttt{0812.0351}

\bibitem{Armillis:2010qk}
R.~Armillis, C.~Coriano, L.~Delle~Rose, Phys. Rev. D \textbf{82}, 064023
  (2010), \texttt{1005.4173}

\bibitem{Armillis:2010ru}
R.~Armillis, C.~Coriano, L.~Delle~Rose, AIP Conf. Proc. \textbf{1317}, 185
  (2010), \texttt{1007.2141}

\bibitem{Coriano:2020ees}
C.~Corian\`o, M.M. Maglio, Phys. Rept. \textbf{952}, 1 (2022),
  \texttt{2005.06873}

\bibitem{Coriano:2012wp}
C.~Coriano, L.~Delle~Rose, E.~Mottola, M.~Serino, JHEP \textbf{08}, 147 (2012),
  \texttt{1203.1339}

\bibitem{Chernodub:2019tsx}
M.N. Chernodub, C.~Corian\`o, M.M. Maglio, Phys. Lett. B \textbf{802}, 135236
  (2020), \texttt{1910.13727}

\bibitem{Chernodub:2017jcp}
M.N. Chernodub, A.~Cortijo, M.A.H. Vozmediano, Phys. Rev. Lett. \textbf{120},
  206601 (2018), \texttt{1712.05386}

\bibitem{Chernodub:2021nff}
M.N. Chernodub, Y.~Ferreiros, A.G. Grushin, K.~Landsteiner, M.A.H. Vozmediano,
  Phys. Rept. \textbf{977}, 1 (2022), \texttt{2110.05471}

\bibitem{bms}
A.~Bzowski, P.~McFadden, K.~Skenderis, Journal of High Energy Physics
  \textbf{2014} (2014)

\bibitem{Bzowski:2017poo}
A.~Bzowski, P.~McFadden, K.~Skenderis, JHEP \textbf{11}, 153 (2018),
  \texttt{1711.09105}

\bibitem{Coriano:2013jba}
C.~Coriano, L.~Delle~Rose, E.~Mottola, M.~Serino, JHEP \textbf{07}, 011 (2013),
  \texttt{1304.6944}

\bibitem{Bzowski:2015pba}
A.~Bzowski, P.~McFadden, K.~Skenderis, JHEP \textbf{03}, 066 (2016),
  \texttt{1510.08442}

\bibitem{Coriano:2019sth}
C.~Corian\`o, M.M. Maglio, JHEP \textbf{09}, 107 (2019), \texttt{1903.05047}

\bibitem{lovelock}
D.~Lovelock, Mathematical Proceedings of the Cambridge Philosophical Society
  \textbf{68}, 345  (1970)

\bibitem{edgar}
S.B. Edgar, A.~Hoglund, J. Math. Phys. \textbf{43}, 659 (2002),
  \texttt{gr-qc/0105066}

\bibitem{Coriano:2019nkw}
C.~Corian\`o, M.M. Maglio, D.~Theofilopoulos, Eur. Phys. J. C \textbf{80}, 540
  (2020), \texttt{1912.01907}

\bibitem{ttjj}
C.~Corian\`o, M.M. Maglio, R.~Tommasi, Work in preparation  (2022)

\end{thebibliography}

\end{document}